\UseRawInputEncoding 
\documentclass[%
reprint,
nofootinbib,
 amsmath,amssymb,
 aps,
]{revtex4-1}

\usepackage{graphicx}
\usepackage{hyperref}
\usepackage{dcolumn}
\usepackage{bm}
\usepackage{braket}
\usepackage{graphicx}
\usepackage{xcolor}
\usepackage[bottom]{footmisc}

\begin{document}

\preprint{APS/123-QED}

\title{Experimentally constrained wave function method}

\author{Stasis Chuchurka}%
\email{stasis.chuchurka@desy.de}
\affiliation{
Deutsches Elektronen-Synchrotron DESY, Hamburg 22607, Germany\;
\\
Department of Physics, Universit\"at Hamburg, Hamburg 22607, Germany
}%

\author{Milaim Kas}%
\affiliation{
Deutsches Elektronen-Synchrotron DESY, Hamburg 22607, Germany\;
\\
Department of Physics, Universit\"at Hamburg, Hamburg 22607, Germany
}%

\author{Andrei Benediktovitch}
\affiliation{
Deutsches Elektronen-Synchrotron DESY, Hamburg 22607, Germany
}%

\author{Nina Rohringer}
\affiliation{
Deutsches Elektronen-Synchrotron DESY, Hamburg 22607, Germany\,
\\
Department of Physics, Universit\"at Hamburg, Hamburg 22607, Germany
}%

\begin{abstract}
In this work, we extend the x-ray constrained wavefunction fitting approach, a key method in quantum crystallography for charge density reconstruction, to incorporate experimental observables beyond x-ray diffraction. Unlike traditional quantum crystallography methods, which are typically limited to molecules in their ground states, our approach integrates excited states. This advancement will enable simultaneous fitting of x-ray diffraction data alongside optical and x-ray spectroscopic data. We introduce a comprehensive theoretical framework that allows for the inclusion of any experimental observable as a constraint in wavefunction reconstruction. Furthermore, we provide detailed derivations and instructions for implementation of this method using two electronic-structure methods: a generalized Hartree-Fock method for excited states and the Coupled Cluster Equation-of-motion method.
\end{abstract}

\maketitle


\section{\label{sec:intro}Introduction}

In the Schr\"odinger picture of quantum mechanics, the wavefunction is the fundamental entity of a quantum system, encapsulating all its properties. However, for all but the simplest systems, obtaining an exact wavefunction is impractical, necessitating the use of approximations. Quantum Chemistry focuses on developing theoretical models to describe molecular systems. These models become increasingly complex and computationally demanding as they strive towards an exact solution. Consequently, their predictive power often becomes limited even for medium-sized molecular systems composed of merely a few tens of atoms.

The atomic and electronic structures of molecular systems can also be derived from experimental measurements. For instance, nuclear magnetic resonance (NMR) spectroscopy \cite{Kwan2011, Bothwell2010, Sels2020} is routinely employed to obtain detailed insights into molecular structures. Cryo-electron microscopy has become a mainstream technique, offering near-atomic resolution \cite{Milne2012, doi:10.1146/annurev-biochem-060614-034226}. One of the most powerful methods for determining the structure of large molecular complexes is x-ray crystallography \cite{table, Giacovazzo2011, Abola2000}, which can precisely map the three-dimensional positions of atoms within molecules, provided the molecules are arranged in a crystal lattice. When high-quality x-ray or electron diffraction data are available, researchers can move beyond simple independent-atom models to reconstruct the valence-electron density between atoms, thereby gaining insights into chemical bonds and related properties of the system. The Quantum Theory of Atoms In Molecules (QTAIM) \cite{1994'Bader} represents a cutting-edge approach in this domain. It provides valuable information on chemical bonding by performing a topological analysis of the electron density. This includes identifying critical points of the electron density, classifying their rank and signature, and examining the Laplacian and the electron density's Hessian matrix \cite{2018'Zhang, 2014'Jorgensen, 2011'Contreras, 2010'Johnson}.

To reconstruct the valence-electron-density with the necessary spatial resolution, appropriate high-quality x-ray/electron intensity diffraction data have to be analysed based on advanced phase-reconstruction and refinement algorithms, such as the Maximum Entropy Method (MEM)\,\cite{2009'Smaalen} or the Multipole Formalism (MP)\,\cite{1997'Coppens}. Unfortunately, these approaches are not universal \cite{2007'Hofmann}: MEM formalism requires a detailed initial guess (prior) of the charge density \cite{2009'Smaalen} to obtain a suitable electron distribution for the determination of bond properties \cite{2009'Netzel}. MP suffers from over-fitting, non-uniqueness, overestimating polar bond electron densities, poor sensitivity to hydrogen atoms, and other deficiencies \cite{2017'Woinska}.

A better structural refinement and deeper understanding of the electron-density distribution can be gained by involving quantum mechanical methods into the analysis of the x-ray crystallography data. Vice versa, the wavefunction obtained from approximate quantum chemistry methods can be improved by means of biasing electronic structure calculations by experimental data. Constraining quantum chemistry calculations by x-ray diffraction (XRD) data, these approaches are known as quantum crystallography \cite{2017'Grabowsky, 2018'Genoni}. These empirical methods turned out fruitful.  Notably, combining electronic structure calculations and XRD data improved the fidelity of the reconstruction of the position of Hydrogen atoms--that are typically difficult to catch with XRD--with accuracy comparable to the golden standard of neutron diffraction \cite{2016'Woinska}.

The first quantum crystallography approaches have been developed in the 1970-s, using the single-particle reduced electron density matrix and relying on iterative matrix equations with XRD data as external constrains \cite{1972'Clinton, 1989'Aleksandrov}. Nowadays, the most widespread quantum crystallography methods are based on the x-ray constrained wavefunction (XCW)-fitting approach developed by D. Jayatilaka \cite{1998'Jayatilaka}. Its main idea is to use XRD data to refine a simple Hartree-Fock (HF) model. Practically, it boils down to recasting the traditional variational principle for the HF wave function into a minimization problem that involves the experimental constraints. The function $L(\bm{\varepsilon},\textbf{c},\lambda)$, given by
\begin{equation}
    \label{minimization problem for the ground state}
    L(\bm{\varepsilon},\textbf{c},\lambda)=E(\bm{\varepsilon},\textbf{c})+\lambda\chi^2(\textbf{c}),
\end{equation}
is being minized. It consists of the HF ground-state energy expression $E(\bm{\varepsilon},\textbf{c})$
and an appropriately weighted measure of the fidelity $\chi^2$. The latter depends on the difference between the predicted $F_{\text{calc.}}(\textbf{h}, \textbf{c})$ and observed $F_{\text{exp.}}(\textbf{h})$ scattering intensities at reciprocal lattice vector $\textbf{h}$
\begin{equation*}
    \chi^2(\textbf{c}) = \frac{1}{N-1}\sum_{\textbf{h}}^N\frac{|F_{\text{exp.}}(\textbf{h})-F_{\text{calc.}}(\textbf{h}, \textbf{c})|^2}{\sigma^2(\textbf{h})},
\end{equation*}
where $N$ is the number of measurements and $\sigma(\textbf{h})$ is the measurement error corresponding to each structure factor. The variational parameters $\textbf{c}$ and $\bm{\varepsilon}$ determine the HF state and are the expansion coefficients to an appropriately chosen single-particle basis expansion of the Slater determinant. The Lagrange multiplier $\lambda$ regulates the balance between theoretical predictions and experimental data in the minimization problem. As $\lambda$ is increased, the solution prioritizes a closer alignment with the experimental data. The optimization of the HF energy $E(\bm{\varepsilon},\textbf{c})$ yields the Roothaan-Hall equations for the expansion coefficients $\textbf{c}$ and orbital energies $\bm{\varepsilon}$
\begin{equation*}
    \textbf{f}\textbf{c}=\textbf{S}\textbf{c}\bm{\varepsilon},
\end{equation*}
where  $\textbf{f}$ is the Fock operator and \textbf{S} represents the overlap matrix of the generally non-orthonormal basis functions. Employing a refined variational principle, as described in Equ.\ (\ref{minimization problem for the ground state}) gives a modified Fock equation 
\begin{equation*}
    \textbf{f} \quad \to \quad \textbf{f}\,\,+\,\textbf{V}_\text{exp.},
\end{equation*}
where $\textbf{V}_\text{exp.}$ is determined by the fidelity term $-\lambda\chi^2(\textbf{c})$ and plays the role of an effective potential, pushing electrons to the positions that reproduce the experimental data. 

The XWR method thus results in a self-consistent system of equations for the variables \textbf{c}, reducing the variational space by joint minimization of the HF energy expression $E(\textbf{c})$ and the cost function conditioned by experimental data. In addition to the ground-state density, the XWR gives an estimate of the ground-state wave function and thereby gives access to ground-state many-particle observables.  

Current XCW implementations are mostly focused on molecular crystals \cite{2017'Grabowsky, 2018'Genoni}. Several commonly used electronic structure methods have been applied and tested within XCW in order to model the corresponding wave function:  Restricted HF formalism \cite{1998'Jayatilaka, 2001'Jayatilaka}, the Density Functional Theory \cite{2012'Jayatilaka}, Extremely Localized Molecular Orbitals Method \cite{2014'DosSantos, 2013'Genoni}. In bench-mark studies, these methods were shown to partially capture electron correlation effects \cite{2017'Genoni_ecorr}. Recently, they have been generalized to include relativistic effects \cite{2016'Bucinsky} and have been extended to a multi-Slater determinant description \cite{2017'Genoni}. 

The full power of quantum crystallography can be achieved by a combination of XCW and the Hirshfeld Atom Refinement (HAR) approach \cite{2014'Capelli} -- the x-ray wavefunction refinement (XWR) approach. XWR's software realization is known under the name Tonto \cite{2003'TONTO}). A thorough validation of XWR \cite{2017'Woinska} demonstrates that XWR clearly outperforms the MP method in reconstructing valence electron densities. The power of XCW to reconstruct accurate wave functions encompassing solid state effects was demonstrated in \cite{2013'Cole}.  Beyond the ground-state densities, this method successfully predicts other ground-state properties: The authors used XCW to calculate optical properties of zinc [tris]thiourea sulphate: the refractive index showed good agreement with the experimental data, in contrast to gas-phase ab initio calculations.

Inspired by the success of quantum crystallography, we propose expanding this concept to incorporate experimental observables beyond x-ray diffraction (XRD). Our goal is to develop a unified electronic structure model that integrates complementary experimental methods, reflecting both ground- and excited-state properties of the underlying quantum mechanical system. Specifically, we seek to create a comprehensive electron-structure model capable of interpreting and fitting XRD data alongside optical and x-ray spectroscopic data. The integration of experimental data from optical and x-ray absorption or emission spectroscopy, including x-ray absorption near edge structure (XANES) and x-ray absorption fine structure (EXAFS) measurements, could establish a robust and versatile approach.

The complementary analysis of x-ray diffraction (XRD) and x-ray emission spectroscopy (XES) is highly valuable for examining (bio)catalytic processes with precision at electronic length and timescales. Groundbreaking experiments conducted at storage-ring based x-ray sources and with x-ray free electron lasers (XFELs) have demonstrated the feasibility of combining XRD and XES in a single, high-quality experiment \cite{Kern2014, doi:10.1021/acs.jpcb.5b12471}. While XRD provides detailed access to the structural arrangement (atomic positions) of a chemical compound or reaction unit, XES serves as a complementary and sensitive probe of changes in chemical bonds. XES offers site-specific information, being sensitive to the local electronic structure and bonding of the absorbing atom \cite{2009'Bergmann}. Valence-to-core XES, in particular, is recognized for its ability to yield insights into chemical bonding \cite{2014'Gallo}. Moreover, XES has been shown to provide information about even the OH bond \cite{2002'Guo, 2005'Odelius}. Like all spectroscopic techniques, XES relies on comparing experimental data to theoretically calculated spectra and cannot be directly inferred from XES data alone. Typically, the atomic model used to fit XRD data is validated by XES measurements. However, a coherent electronic structure model that simultaneously and self-consistently incorporates both complementary datasets is lacking. We are targeting the development of such a unified model. This approach, inspired by the principles of x-ray constrained wavefunction (XCW) fitting, would enable the reconstruction of electronic ground-state properties and the establishment of a reliable electronic structure model that is also responsive to excited-state properties.

In contrast to XRD, the theoretical characterization of XES involves information of excited-state wave functions, such as transition-dipole moments and excited-state energies. An XCW approach based on the ground-state HF equations is thus inappropriate. A simple inclusion of a weighted $\chi^2$ statistic comparing predicted and observed XES spectrum in equation (\ref{minimization problem for the ground state}) generates equations only for the ground state variables, and is thus insufficient.

In Sec.~\hyperref[sec:II]{II}, we extend the main idea of quantum crystallography to include many-level systems and treat ground- and excited-state properties on the same footing. It formally allows for the inclusion of any experimental observable as a constraint for the wave function reconstruction. In this conceptual consideration, we do not perform any particular expansion of the many-particle wave function. Similarly to the x-ray constrained  wave function(XCW)-fitting approach, the resulting method based on our new equations is called Experimentally Constrained Wavefunction(ECW)-fitting.

For future numerical applications, we suggest and consider two practically relevant parametrizations of the many-body wavefunction: The HF method \cite{Szabo} generalized to excited states \cite{Hardikar2020a, Gilbert2008} and the Coupled Cluster (CC) Theory \cite{Bartlett_book, Bartlett2007} in the framework of the  Equation-of-motion (EOM) method \cite{Stanton1993}.

In Sec.~\hyperref[sec:HF method]{III}, we start with a less complex unconstrained HF ansatz giving a relatively compact system of equations but expected low level of accuracy. In practice, the stability of the equations for excited states might become an issue: solutions for excited states may jump to the global minimum during the optimization iterations. Besides, the single-determinant unconstrained ansatz frequently causes the so-called spin contamination. Fortunately, these problems have been already addressed \cite{Gilbert2008, Andrews1991}, and  the proposed techniques can be applied to our formalism.

In Sec.~\hyperref[sec:III]{IV}
we formulate the novel ECW-fitting approach in terms of the Coupled Cluster method and give a detailed derivation of the resulting equations of the experimentally-constrained Coupled Cluster equations (EC-CC). CC method -- often considered as the golden standard of quantum chemistry methods \cite{Helgaker2014} -- is size-extensive and frequently used to make highly accurate electronic structure calculations. Together with the Equation-of-motion method \cite{Stanton1993} it becomes a powerful tool for the analysis of the excited states. Besides that, the  ground and excited  states  are  well distinguished  and  have  separate parametrization, allowing more freedom for fitting experimental data related to different states. The ab initio quantum chemistry program Q-Chem \cite{qChem} contains a diverse set of the methods for studying electronically excited states, including Coupled Cluster-Equation of Motion (CC-EOM). The derived equations are therefore well suited to serve as a basis for implementing EC-CC within the Q-Chem software.

\section{The experimentally Constrained Wavefunction (ECW) Approach}
\label{sec:II}

If spectroscopic or any other experimental data beyond XRD is incorporated in a quantum crystallography frame, one has to account not only for the ground state $\ket{\Psi_0}$ of the system, but also for a set of excited states $\ket{\Psi_1},\ket{\Psi_2},\,...$. In the traditional XCW-fitting approach, the ground-state electronic density is used to predict the electronic structure factors $F_{\text{exp.}}$. Spectroscopic data generally  involve transition energies $E^{nm}_{\text{exp.}}$ and transition-dipole moments $\textbf{d}^{nm}_{\text{exp.}}$ and encode excited-state properties:
\begin{equation*}
    \textbf{d}^{nm}_{\text{calc.}}=\braket{\Psi_n|\hat{\textbf{d}}|\Psi_m}, \, E^{nm}_{\text{calc.}}=\braket{\Psi_m|\hat{K}|\Psi_m}-\braket{\Psi_n|\hat{K}|\Psi_n}
\end{equation*}
Here, $\hat{K}$ is the kinetic energy operator\footnote{According to virial theorem, the total energy of non-relativistic electronic system is equal to the negative of its kinetic energy.}. Note, that the involved properties correspond to one-particle operators, that posses a convenient form after second quantization
\begin{equation}
\label{spectroscopy operators}
    \hat{\textbf{d}}=\sum_{pq}\textbf{d}_{pq}\hat{a}_p^\dag\hat{a}_q,\quad \hat{K}=\sum_{pq}K_{pq}\hat{a}_p^\dag\hat{a}_q.
\end{equation}
Here, $\hat{a}_p^\dag$ and $\hat{a}_q$ are fermionic operators, creating and annihilating electrons in one-particle spin-orbitals $\ket{\phi_p}$, $\ket{\phi_q}$ of a conveniently chosen single-particle representation of the electronic field modes (usually atomic or molecular orbitals). The operator matrix-elements
\begin{equation*}
    \textbf{d}_{pq}=\braket{\phi_p|\textbf{d}^{(0)}|\phi_q}, \quad K_{pq}=\braket{\phi_p|K^{(0)}|\phi_q}
\end{equation*}
are known beforehand \footnote{The superindex $^{(0)}$ indicates that the operator only corresponds to one particle}. The expectation values are conveniently represented by the one-particle reduced (transition) density matrix $\gamma^{nm}$
\begin{equation}
    \label{rdm1}
    \gamma^{nm}_{pq}=\braket{\Psi_m|\hat{a}_q^\dag\hat{a}_p|\Psi_n}.
\end{equation}
Its matrix elements relate physical many-particle states $\ket{\Psi_n}$ and $\ket{\Psi_m}$ to predefined one-particle spin-orbitals $\ket{\phi_p}$, $\ket{\phi_q}$. The observables are represented in terms of $\gamma^{nm}_{pq}$ by 
\begin{equation*}
    \textbf{d}^{nm}_{\text{calc.}}=\sum_{pq}\textbf{d}_{pq}\gamma^{mn}_{qp},\quad E^{nm}_{\text{calc.}}=\sum_{pq}K_{pq}\left(\gamma^{mm}_{qp}-\gamma^{nn}_{qp}\right).
\end{equation*}
A generalization to other one-particle observables $\hat{A}$ is straightforward. If a measured property  $A^{nm}_{\text{exp.}}$ is connected to a transition matrix element of the operator  $A^{nm}:=\bra{\Psi_n}\hat{A} \ket{\Psi_m}$  its predicted value $A_{\text{calc.}}^{nm}$ reads 
\begin{equation}
\label{A through gamma}
    A^{nm}_{\text{calc.}}=\sum_{pq}A_{pq}\gamma^{mn}_{qp}, \quad A_{pq}=\braket{\phi_p|A^{(0)}|\phi_q}.
\end{equation}

To compare measured and calculated properties, the  XCW-fitting approach suggests using an additional term in the variational problem (\ref{minimization problem for the ground state}) that quantifies the reliability (loss or cost function) of the theoretically determined quantity to reproduce the experimental data. The loss-function is often chosen as a mean-square $L^2$ deviation $\chi^2$, but this is only one possible choice, typically motivated by the experimental noise being assumed following Gaussian statistics. If the experimental data is contaminated by the impulse noise, $L^1$ norm would be more appropriate as cost function \cite{Bar2006}. Compared to the $L^2$ norm, $L^1$ is more robust against mixed or unknown noise \cite{doi:10.1137/130904533}. Besides that, we do not have to restrain ourselves from adding additional constraints, based on experimental or mathematical considerations\footnote{For example, one can add a term that minimizes the overlap of the different electronic states, that due to the experimental constraints are not necessarily orthonormal with respect to each other.}. Irrespective of the noise type, the $L^1$ norm can be used to force insignificant parameters to disappear and is often used in regularization schemes. This has been, for example, successfully implemented within the Coupled Cluster approach \cite{Ivanov2017}. 

Finally, we collect all the constraints, multiplied by  separate Lagrange multipliers, into one new function $Q[\Psi_0,\Psi_1,..., \Psi_N]$
\begin{equation}
    \label{equation for Q}
    \lambda\chi^2(\bm{\varepsilon},\textbf{c})\quad \to \quad Q[\Psi_0,\Psi_1,..., \Psi_N]
\end{equation}
The structure factors $F_{\text{exp.}}$ from traditional XCW are replaced by a set of measured values $A^{nm}_{j,\text{exp.}}$, where $j$ stands for different properties. Since our formalism involves both ground and excited states, we formulate $N$ coupled minimization problems
\begin{equation}
\label{general minimization problem}
\begin{split}
\min_{\,\,\,\Psi_n}L_n,\quad\quad\quad\quad\quad\quad\quad\\  L_n\!=\!\bra{\Psi_n}\hat{H}-E_{n}\ket{\Psi_n}+\,Q[\Psi_0,\Psi_1,..., \Psi_N].
\end{split}
\end{equation}
Here, we minimize each function $L_n$ only with respect to $\ket{\Psi_n}$. The other states are considered frozen, but obtained from the remaining minimization problems in a self-consistent manner. The self-consistent treatment of the variational problem (\ref{general minimization problem}) thus gives $N$ equations for $N$ states $\Psi_n$. In the absence of $Q[\Psi_0,\Psi_1,...]$, an excited state corresponds to a local minimum of the variational problem. Experimental constraints related to the states $\ket{\Psi_i}$ transform the local minimum for states $\ket{\Psi_i}$ into a global one (up to degeneracy).

The variational problem expressed in equation (\ref{general minimization problem}) can be recast into a set of nonlinear Schr\"odinger-like equations
\begin{equation}
\label{Sch_gene}
    \hat{H}\ket{\Psi_n}+\sum_m\hat{V}_{\text{exp.}}^{nm}\ket{\Psi_m}=E_{n}\ket{\Psi_n}.
\end{equation}
The resulting states $\ket{\Psi_n}$ correspond to experimentally constrained wave functions. Hereafter, we will refer to equations (\ref{Sch_gene}) as the Experimentally Constrained Schr\"odinger Equations -- a foundation of our experimentally constrained wavefunction(ECW)-fitting approach.
Constraint function $Q$ is thus mapped into an effective one-electron potential operator $\hat{V}_{\text{exp.}}$ given by
\begin{eqnarray}
\label{experimental_potential}
        \hat{V}^{nm}_{\text{exp.}}=\sum_j\frac{\partial Q}{\partial A^{nm}_{j}}\hat{A}_j.
\end{eqnarray}
This operator can be interpreted as a potential that pushes the electrons towards the positions that reproduce the experimental properties, and is seen as a perturbation of the underlying theoretical model determined by its Hamiltonian $\hat{H}$. 

Starting from our general formulation, any kind of quantum chemistry method could be applied to equation (\ref{general minimization problem}) or (\ref{Sch_gene}), by specifying a distinct form of the Hamiltonian and the structure of the wave function.  We have decided to use the HF and Coupled Cluster theories as the underlying models for the description of the wavefunctions $\Psi_n$. 

\section{Hartree-Fock method}
\label{sec:HF method}

The Hartree-Fock method is a basic electronic structure approximation based on single-particle orbitals. Due to the simplest parametrization of the electronic wavefunction, it treats electron correlations only approximately. Slightly modifying HF methods, one can construct more practical exact-exchange Density Functional Theories (DFT) \cite{Koch2001} showing performance comparable to high-order post-Hartree-Fock methods \cite{doi:10.1021/acs.jctc.0c01082}. Following the initial steps of the XCW method, we formulate our theory in terms of the basic Hartree-Fock method adopted for excited states \cite{Hardikar2020a, Gilbert2008}, which is already expected to improve the quality of the computed observables. The extension to Density Functional Theory is straightforward and only requires modification of the exchange potential.

For ground state $\Psi_0$ and each excited state $\Psi_n$, we introduce spin-unconstrained independent Slater determinants
\begin{equation*}
    \ket{\Psi_n}=\frac{1}{\sqrt{N!}}\det{\left[\ket{\chi_{1}^{({n})}}\ket{\chi_{2}^{({n})}}\ket{\chi_{3}^{({n})}}...\right]}
\end{equation*}
with an independent set of $N$ spin-orbitals $\ket{\chi_{a}^{({n})}}$ for each state $\Psi_n$. Here, $N$ is the number of electrons. The spin-orbitals $\ket{\chi_{a}^{({n})}}$ can be decomposed in a common non-orthogonal basis $\left\{\ket{\chi_{\nu}}\right\}$ 
\begin{equation*}
    \ket{\chi_{a}^{({n})}}= \ket{\chi_{\nu}}C_{\nu a}^{(n)}.
\end{equation*}
Note that we use Greek letters for the global common basis. Generally speaking, they basis set may be non-orthogonal 
\begin{equation*}
    S_{\mu\nu}=\braket{\chi_{\mu}|\chi_{\nu}}.
\end{equation*}
Although the spin-orbitals within one state must be orthogonalized, the spin-orbitals between different states can be generally non-orthogonal. Using expansion coefficients $C_{\nu a}^{(n)}$ as a transformation matrix, $S_{\mu\nu}$ turn into the overlaps between the spin-orbitals
\begin{equation*}
     \braket{\chi_{a}^{({n})}|\chi_{b}^{({m})}}=\Sigma^{nm}_{ab}=\sum_{\mu,\nu}C_{\mu a}^{(n)*}S_{\mu\nu}C_{\nu b}^{(m)}.
\end{equation*}
Later, these overlaps appear frequently so we introduce special notation $\Sigma^{(nm)}_{ab}$ for them. For example, the determinant of the matrix $\Sigma^{(nm)}$ constructed out of $\Sigma^{(nm)}_{ab}$ is the overlap between ground and excited many-body states $\Psi_n$ and $\Psi_m$
\begin{equation}
    \label{eq: overlap between many-body states}
    \braket{\Psi_n|\Psi_m}=\text{det}\,\Sigma^{nm}.\ 
\end{equation}
Since the ground and excited states have independent parametrization, they may be generally non-orthogonal \cite{Gilbert2008}. This overlap can be included in constrain function $Q$ to ensure the orthogonality.

Before we derive the equations for the expansion coefficients $C_{\nu a}^{(n)}$, let us show how they can be used to calculate one-particle reduced density matrices (\ref{rdm1}) required for constructing one-particle observables (\ref{A through gamma}). Reduced density matrices (\ref{rdm1}) can be seen as overlaps between pairs of ionized states. It means that we can go back to determinant (\ref{eq: overlap between many-body states}) and take out the lines of matrix $\Sigma^{(nm)}$ corresponding to the annihilated electrons. Since electrons are fermions, we also have to take into account the change of the sign. In mathematics, all these operations are embodied in the notion of the cofactor
\begin{equation*}
    \gamma^{mn}_{ab}=\braket{\Psi_n|\hat{a}_b^{(n)\dag}\hat{a}_a^{(m)}|\Psi_m}=\text{Cofactor}_{b,a}\,\Sigma^{nm},
\end{equation*}
where $a$ and $b$ specify the row and column to be taken out. Note that the creation and annihilation operators require indices $n$ and $m$ since different states have independent spin-orbitals. To make use of $\gamma^{mn}_{ab}$, spin-orbitals $\bra{\phi_p}$, $\ket{\phi_q}$ in equation (\ref{A through gamma}) must consequently correspond to states $\Psi_n$ and $\Psi_m$ respectively. In order to prevent confusion, it makes sense to express density matrix elements $\gamma^{nm}_{pq}$ using the common basis set
\begin{equation}
    \label{eq: one particle gammas}
   \gamma^{mn}_{\mu\nu}=\sum_{a,b}C_{\mu a}^{(m)}\gamma^{mn}_{ab}C_{\nu b}^{(n)*}.
\end{equation}
Then, $\bra{\phi_p}$, $\ket{\phi_q}$ in equation (\ref{A through gamma}) both are the spin-orbitals of the common basis set. To distinguish matrix elements in the common basis set, we use Greek indices. If we assume $n=m$, the cofactors disappear and we get well-known expression $\gamma^{nn}_{\mu\nu}=\sum_a C_{\mu a}^{(n)}C_{\nu a}^{(n)*}$.

In addition to the one-particle matrices, it is convenient to introduce two-particle reduced density matrices to represent the equations for the expansion coefficients in the most compact form
\begin{equation*}
\gamma^{mn}_{aba'b'}=\bra{\Psi_n}\hat{a}^{(n)\dag}_{a'}\hat{a}^{(n)\dag}_{b'}\hat{a}^{(m)}_{b}\hat{a}^{(m)}_{a}\ket{\Psi_m}.
\end{equation*}
Since they involve two pairs of creation and annihilation operators, it effectively corresponds to double ionization of the initial states, so we can simply use a modified notion of cofactor
\begin{equation*}
\gamma^{mn}_{aba'b'}=\text{Cofactor}_{a'b',ab}\,\Sigma^{nm},
\end{equation*}
where $a'$ and $b'$ are the rows and $a$ and $b$ are the columns to be taken out. In the common basis set, we have
\begin{equation}
\label{eq: two particle gammas}
\gamma^{mn}_{\mu\nu\mu'\nu'}=\sum_{a,b,a',b'}C_{\mu a}^{(m)}C_{\nu b}^{(m)}\gamma^{mn}_{aba'b'}C_{\mu'a' }^{(n)*}C_{\nu'b'}^{(n)*}.
\end{equation}

There are multiple ways to generate the equations for expansion coefficients $C_{\nu p}^{(n)}$. Starting from equation (\ref{Sch_gene}), the most straightforward way is to apply Brillouin's theorem. First, we prepare a singly excited state $\hat{a}^{(n)\dag}_i\hat{a}^{(n)}_a\ket{\Psi_n}$. Indices $i,j,k,l...$ stand for the virtual orbitals, whereas $a,b,c,d...$ correspond to the occupied ones. Now, we multiply equation (\ref{Sch_gene}) by the resulting excited state 
\begin{equation*}
\label{Brillouin}
    \bra{\Psi_n}\hat{a}^{(n)\dag}_a\hat{a}^{(n)}_i\left[\hat{H}\ket{\Psi_n}+\sum_m\hat{V}_{\text{exp.}}^{nm}\ket{\Psi_m}\right]=0,
\end{equation*}
which gives a set of independent equations for different combinations of indices $a$, $i$ and $n$. The first term generates matrix elements  $f_{ia}^{(n)}$ of the well-known Fock operator for state $n$. The rest comes from the experimental potentials that we split into two contributions
\begin{equation}
\label{modified fock}
    \bar{f}_{ia}^{(n)}= f_{ia}^{(n)}+v_{ia}^{(n)}+\sum_{m\neq n} v_{ia}^{(n,m)}=0,
\end{equation}
where  $v_{ia}^{(n)}$ takes into account the effect of the experimental data related solely to state $n$
\begin{equation*}
    v_{ia}^{n}=\sum_{\sigma,
    \nu}C_{\sigma i}^{(n)*}V^{nn}_{\sigma\nu}C_{\nu a}^{(n)},
\end{equation*}
and $v_{ia}^{(n,m)}$ represents the experimental data related to the pairs of different states
\begin{multline*}
    v_{ia}^{nm}=\sum_{\sigma,
    \delta,\nu,\mu}C_{\sigma i}^{(n)*}\big[ V^{nm}_{\sigma\nu}\gamma^{mn}_{\nu\mu}\\+S_{\sigma \nu}\sum_{\nu',\,\mu'}\gamma^{mn}_{\nu\mu'\mu\nu'}V^{nm}_{\nu'\mu'}\big]S_{\mu\delta}C_{\delta a}^{(n)}.
\end{multline*}

Now, we have to find an orthogonal system of virtual and occupied spin-orbitals that would satisfy the system of equations (\ref{modified fock}). Although this system contains enough equations to separate the space into the occupied and virtual subspaces, it would be convenient to generate more equations and to define all the spin-orbitals up to insignificant phase-multipliers. To get the occupied-virtual quadrant of the modified Fock matrix $\bar{f}_{ai}^{(n)}$, we have to simply mirror and conjugate the virtual-occupied quadrant $\bar{f}_{ai}^{(n)}=\bar{f}_{ia}^{(n)*}$. To get the occupied-occupied and virtual-virtual quadrants we have full freedom with the only condition: the Fock matrix must be hermitian. For example, we could extrapolate the equation (\ref{modified fock}). As shown in \cite{1998'Jayatilaka}, a better alternative can be to use the traditional Fock matrix for the occupied-occupied and virtual-virtual quadrants without any experimental contributions $\bar{f}_{ab}^{(n)}=f_{ab}^{(n)}, \bar{f}_{ij}^{(n)}=f_{ij}^{(n)}$. The problem becomes singular when constraint function $Q$ is large since there is more parameters than data. Consequently, it is more preferable not to create more sources of instability. Finally, we get the following system of equation
\begin{equation}
    \label{eq: Roothaan 1}
    \bar{f}_{pq}^{(n)}=\Sigma^{nn}_{pq}\varepsilon_q.
\end{equation}
Since the spin-orbitals are assumed to be normalized and orthogonal, overlap $\Sigma^{nn}_{pq}$ equals zero whenever $p\neq q$, and equations (\ref{modified fock}) are restored. If $p=q$, we get diagonal elements of the traditional Fock matrix $\bar{f}_{qq}^{(n)}=f_{qq}^{(n)}$ that are typically interpreted as orbital energies $\varepsilon_q$ used for the Aufbau principle. 

Usually, equations (\ref{eq: Roothaan 1}) are formulated in terms of Roothaan equations resembling an eigenvalue problem. Using expansion coefficients as a transformation matrix and multiplying equation (\ref{eq: Roothaan 1}) by it, one gets
\begin{equation}
    \label{eq: Roothaan 2}
    \sum_\nu \left(\bar{f}_{\mu\nu}^{(n)}-\varepsilon_q S_{\mu\nu}\right)C_{\nu q}^{(n)}=0.
\end{equation}
The expression for $\bar{f}_{\mu\nu}^{(n)}$ is too bulky and technical to be mentioned in the main part of the article, so we leave it in Appendix \ref{technical}.

Before solving the equations, we define the relative importance of theory and experiment by setting specific values for the Lagrange multipliers in the experimental constraint $Q[\Psi_0,\Psi_1,..., \Psi_N]$. Equations (\ref{eq: Roothaan 2}) are typically solved iteratively. To summarize, the steps needed to perform a numerical investigation of a molecular system are the following:

\begin{enumerate}
\setlength\itemsep{1em}
\item{We start with a guess that can be taken from a usual Hartree-Fock calculation. For the excited states, our starting guess is a single particle excitation of the ground-state guess.}

\item{We compute one- and two- particle reduced density matrix elements (\ref{eq: one particle gammas}) and (\ref{eq: two particle gammas}). Then, we use them to construct the experimental potentials and  the modified Fock matrix (\ref{eq: technical}).}

\item{We interpret equations (\ref{eq: Roothaan 2}) as an eigenvalue problem. To identify the occupied orbitals among the eigenvectors, we modify the Aufbau principle in accordance with \cite{Gilbert2008}. We choose the eigenvectors that have maximum overlap with the occupied orbitals from the previous iteration.}

\item{Go to step number two unless the convergence criteria are satisfied. }
\end{enumerate}

This iterative procedure can be augmented by improved iterative schemes such as DIIS \cite{PULAY1980393} and its modifications \cite{doi:10.1063/1.1470195, doi:10.1063/1.3304922}. This, however, does not change the presented recipe conceptually. 

\section{Coupled Cluster theory}
\label{sec:III}

A complete description of the Coupled Cluster (CC) theory is beyond the scope of this paper, additional information can be found in the literature \cite{Bartlett_book, Bartlett2007}. Beside its popularity and efficiency at recovering a larger part of the electron correlation, the main reason for this choice is the specific structure of the theory, in which the ground state and excited states are well distinguished and have separate parametrization, allowing more freedom for fitting experimental data related to different states. In this section, we provide the reader with a brief overview of CC and Equation-of-motion (EOM) method. If CC method is not required, one can still adopt the resulting EOM equations for the configuration interaction calculations. 

\subsection{Introduction to Coupled Cluster theory}
\label{sec:III A}
In this section, we consider a traditional Schr\"odinger equation $\hat{H}\ket{\Psi_0}=E_0\ket{\Psi_0}$. Later, we will apply the formalism to our Experimentally Constrained Schr\"odinger equations (\ref{Sch_gene}).  The CC ground state is obtained using the following ansatz of a non-unitary transformation of the HF ground state  {\cite{CC_original}}:
\begin{equation}
\label{GS_CC}
\ket{\Psi_{0}}=Ce^{\hat{T}}\ket{\psi_{\text{HF}}}
\end{equation}           
where $C$ is a normalization constant and $\psi_{HF}$ is the HF single determinant built out of one-particle orbitals $\ket{\phi_p}$. $\hat{T}=\hat{T}_{1}+\hat{T}_{2}+...$ is an excitation operator that performs single, double, ... excitation
\begin{equation*}
    \hat{T}_1 = \sum_{i,a} t^i_a \hat{a}_i^\dag\hat{a}_a, \quad  \hat{T}_2 = \frac{1}{4}\sum_{i,j,a,b} t^{ij}_{ab} \hat{a}_i^\dag\hat{a}_j^\dag\hat{a}_b\hat{a}_a, \quad\dots
\end{equation*}
Here, indices $i,j,k,l...$ stand for the virtual orbitals, whereas $a,b,c,d...$ correspond to the occupied ones. The operator $\hat{T}_n$ has the general form
\begin{equation*}
    \hat{T}_n = \frac{1}{n!^2}\!\!\sum_{i,j,a,b...}\!\! t^{ij...}_{ab...} \hat{a}_i^\dag\hat{a}_j^\dag...\,\hat{a}_b\hat{a}_a.
\end{equation*}
The exponential $e^{\hat{T}}$ is never explicitly expanded for the calculation of the actual ground state (\ref{GS_CC}). Applying the Baker-Campbell-Hausdorff equation,  the ground state properties (\ref{rdm1_GS}) are directly determined. 

The main idea of CC is to formulate equations for the $t$-amplitudes, introduced in the expressions for $\hat{T}$. To generate these equations, one introduces ansatz (\ref{GS_CC}) in the Schr\"odinger equation and multiplies it by $e^{-\hat{T}}$ from the left
\begin{equation}
\label{half of the way}
    e^{-\hat{T}}\hat{H}e^{\hat{T}}\ket{\psi_{\text{HF}}}=E\ket{\psi_{\text{HF}}}.
\end{equation}
This equation, in turn, is projected onto excited determinants $\bra{\psi_{\nu}}=\bra{\psi_{\text{HF}}}\hat{\nu}$, where $\hat{\nu}$ is a product of creation and annihilation operators, promoting electrons from occupied to virtual orbitals. The projection yields
\begin{equation}
\label{T_eq}
    \bra{\psi_{\nu}}\bar{H}\ket{\psi_{HF}} = 0, \quad \bar{H}=e^{-\hat{T}}\hat{H}e^{\hat{T}},
\end{equation}
where $\bar{H}$ is the similarity transformed Hamiltonian. The resulting equations are usually called $T$-equations \cite{Stanton1991}. Different degrees of approximation are obtained with different levels of truncation of $\hat{T}$. The most common level of truncation, $\hat{T}=\hat{T}_{1}+\hat{T}_{2}$, corresponds to the CCSD (Coupled Cluster with single and double particle-hole excitations) method \cite{Stanton1991}. After this truncation, the Baker-Campbell – Hausdorff formula gives a finite expression in terms of $t$-amplitudes for $\bar{H}$, making it convenient for numerical calculations. 
Using the $t$-amplitudes alone, one can get the energy of the ground state by projecting (\ref{half of the way}) onto the HF state $\bra{\psi_{\text{HF}}}$
\begin{equation}
\label{ground state energy}
    E=\bra{\psi_{\text{HF}}}\bar{H}\ket{\psi_{\text{HF}}}.
\end{equation}
To express other ground state properties
\begin{equation*}
    A^{00}_\text{calc.}=\bra{\Psi_0}\hat{A}\ket{\Psi_0}
\end{equation*}
through the $t$-amplitudes, we have to use the CC ansatz (\ref{GS_CC}) again
\begin{equation*}
    A^{00}_\text{calc.}\!=\!|C|^2\bra{\psi_\text{HF}}e^{\hat{T}^\dag}\hat{A}e^{\hat{T}}\ket{\psi_\text{HF}}\!=\!|C|^2\bra{\psi_\text{HF}}e^{\hat{T}^\dag}e^{\hat{T}}\bar{A}\ket{\psi_\text{HF}}
\end{equation*}
In addition to $\bar{H}$, we introduce a similarity transformed $\bar{A}=e^{-\hat{T}}\hat{A}e^{\hat{T}}$ for the sake of consistency. Since $e^{\hat{T}^\dag}\neq e^{-\hat{T}}$, it is a common approach, to introduce a new operator $\hat{\Lambda}$
\begin{equation}
\label{Lambda}
    C^*\bra{\Psi_0}e^{\hat{T}}=|C|^2\bra{\psi_\text{HF}}e^{T^\dag}e^{\hat{T}}=\bra{\psi_\text{HF}}(1+\hat{\Lambda}).
\end{equation}
Typically, $\hat{\Lambda}$ is treated independently from $\hat{T}$, making bra- and ket-vectors unrelated \cite{Salter1989}. This formulation can be retrieved, for example, from response theory.  Since $\bra{\psi_\text{HF}}$ cannot be deexcited, $\hat{\Lambda}$ does not contain excitation operators. For the ground state to be normalized, $\hat{\Lambda}$ must only induce relaxations. Similarly to $\hat{T}$,   $\hat{\Lambda}=\hat{\Lambda}_{1}+\hat{\Lambda}_{2}+...$ has the following decomposition in terms of $\lambda$-amplitudes
\begin{equation*}
    \hat{\Lambda}_1 = \sum_{i,a} \lambda^a_i \hat{a}_a^\dag\hat{a}_i, \quad  \hat{\Lambda}_2 = \frac{1}{4}\sum_{i,j,a,b} \lambda^{ab}_{ij} \hat{a}_a^\dag\hat{a}_b^\dag\hat{a}_j\hat{a}_i, \quad\dots
\end{equation*}
The recipe for (\ref{T_eq}) applied to the bra-vector $\bra{\Psi_0}$ gives
\begin{equation}
\label{Lam_eq}
    \bra{\psi_{HF}}(1+\hat{\Lambda})\bar{H}\ket{\psi_{\nu}} = E\bra{\psi_{HF}}\hat{\Lambda}\ket{\psi_{\nu}}.
\end{equation}
Since $e^{\hat{T}^\dag}\neq e^{-\hat{T}}$, in contrast to (\ref{T_eq}), the resulting equations contain $\hat{\Lambda}$ (\ref{Lambda}). Once the $T$-equations (\ref{T_eq}) are solved, one can proceed to the $\Lambda$-equations (\ref{Lam_eq}). Once the $t$- and the $\lambda$-amplitudes are obtained, one can get the values of the reduced density matrix elements (\ref{rdm1})
\begin{equation}
\label{rdm1_GS}
\gamma_{qp}^{00} = \bra{\psi_{HF}}\hat{\Lambda}e^{-\hat{T}}\hat{a}_{p}^{\dag}\hat{a}_{q}e^{\hat{T}}\ket{\psi_{HF}},
\end{equation}
generating all one-particle properties for the ground state
\begin{equation*}
    A^{00}_{\text{calc.}}=\sum_{pq}A_{pq}\gamma^{00}_{qp}, \quad A_{pq}=\braket{\phi_p|A^{(0)}|\phi_q}.
\end{equation*}

\subsection{Introduction to Equations of Motion theory}
\label{sec: III B}
In this subsection we still consider a traditional Schr\"odinger equation $\hat{H}\ket{\Psi_n}=E_n\ket{\Psi_n}$. Equation of Motions (EOM) theory defines the excited states $\ket{\Psi_n}$ through the ground state $\ket{\Psi_0}$, by promoting the latter with an excitation operator $\hat{R}(n)$
\begin{equation}
\label{ES_CC_R}
    \ket{\Psi_{n}}=\hat{R}(n) \ket{\Psi_0}= Ce^{\hat{T}}\hat{R}(n)\ket{\psi_{HF}},
\end{equation}
where $C$ is the normalization coefficient from the previous subsection. $e^{\hat{T}}$ and $ \hat{R}(n)$ commute, because both operators only excite by definition.  $\hat{R}(n)=r_0(n)+\hat{R}_{1}(n)+\hat{R}_{2}(n)+...$ contains excitations of single, double, ... particle-hole character
\begin{equation*}
    \hat{R}_1(n) \!=\! \sum_{i,a} r^i_a(n) \hat{a}_i^\dag\hat{a}_a, \quad  \hat{R}_2(n) \!=\! \frac{1}{4}\!\sum_{i,j,a,b} r^{ij}_{ab}(n) \hat{a}_i^\dag\hat{a}_j^\dag\hat{a}_b\hat{a}_a,
\end{equation*}
and contains $r_0(n)$ that leaves the state unchanged.
A similar ansatz is introduced for the bra-vectors  $\bra{\Psi_n}$
\begin{equation}
\label{ES_CC_L}
C\bra{\Psi_{n}}=\bra{\psi_{HF}}\hat{L}(n)e^{-\hat{T}},
\end{equation}
where $\hat{L}(n)=\hat{L}_{1}(n)+\hat{L}_{2}(n)+...$ is a deexcitation operator that performs single, double, ... deexcitation
\begin{equation*}
    \hat{L}_1(n)\! =\! \sum_{i,a} l^a_i(n) \hat{a}_a^\dag\hat{a}_i, \quad  \hat{L}_2(n) \!=\! \frac{1}{4}\!\sum_{i,j,a,b} l^{ab}_{ij}(n) \hat{a}_a^\dag\hat{a}_b^\dag\hat{a}_j\hat{a}_i.
\end{equation*}
To make the excited states orthogonal to the ground state, $\hat{L}(n)$ must contain only deexcitation operators. The introduced parametrization for the excited states can be seen as a configuration interaction (CI) expansion based on the correlated ground state \cite{Bartlett2007}. Similarly to $\hat{T}$ and $\hat{\Lambda}$, $\hat{R}(n)$ and $\hat{L}(n)$ are independent and different degrees of approximation are obtained with different levels of their truncation.

The recipe for the $T$-equations (\ref{T_eq}) together with the new ansatz (\ref{ES_CC_R}), (\ref{ES_CC_L}) results in \cite{Bartlett2007}
\begin{equation}
    \begin{split}
        \label{EOM}
   \bra{\psi_{HF}}\bar{H}\hat{R}(n)\ket{\psi_{HF}} &= E_n r_0(n),\\
   \bra{\psi_{\nu}}\bar{H}\hat{R}(n)\ket{\psi_{HF}} &= E_n \bra{\psi_{\nu}}\hat{R}(n)\ket{\psi_{HF}}, \\ 
   \bra{\psi_{HF}}\hat{L}(n)\bar{H}\ket{ \psi_{\nu}} &= E_n \bra{\psi_{HF}}\hat{L}(n)\ket{\psi_{\nu}},
    \end{split}
\end{equation}
In the traditional formulation, $\bar{H}$ depends only on the precalculated $t$-amplitudes, so equations (\ref{EOM}) become a system of linear equations with respect to $r$- and $l$-amplitudes, parametrizing the $\hat{R}(n)$ and the $\hat{L}(n)$ operators.  

After diagonalizing the system of equations (\ref{EOM}) and obtaining the $r$-, $l$-amplitudes and energies, one can calculate the excited states one-electron reduced density matrix elements (\ref{rdm1})
\begin{equation}
\label{rdm1_ES}
    \gamma^{mn}_{qp}=\bra{\psi_{HF}}\hat{L}(n) e^{-\hat{T}}\hat{a}^{\dag}_{p}\hat{a}_{q}e^{\hat{T}}\hat{R}(m)\ket{\psi_{HF}}.
\end{equation}
If transitional properties between the excited and ground state are needed, the same expression can be used with $\hat{L}(0)=1+\hat{\Lambda}$ and $\hat{R}(0)=1$. Finally, the normalization condition implies the following equation
\begin{equation}
    \label{normalization}
    \braket{\Psi_n|\Psi_n}=\bra{\psi_{\text{HF}}}\hat{L}(n)\hat{R}(n)\ket{\psi_{\text{HF}}}=1.
\end{equation}

Before starting the numerical analysis, one needs to recast all the formal equations into an implementable algebraic form. First, one decides on the truncation of $\hat{T}$, $\hat{\Lambda}$, $\hat{R}$ and $\hat{L}$. Using the diagrammatic technique \cite{Crawford2007}, one expresses the terms in the equations directly through the real $t-$, $\lambda-$, $r-$ and $l-$amplitudes.  
To summarize, the steps needed to perform a numerical coupled cluster investigation of a molecular system are the following:
\begin{enumerate}
\setlength\itemsep{1em}
    \item Solve for the $t$ amplitudes using the $T$-equations (\ref{T_eq}) to calculate the energy of the ground state (\ref{ground state energy}).
    \item If properties other than the energy are needed, solve for the $\Lambda$ amplitudes using equation (\ref{Lam_eq}). Use the $t$ and $\Lambda$ amplitudes to construct the reduced density matrices (\ref{rdm1_GS}).
    \item In order to calculate excitation energies, diagonalize system of equations (\ref{EOM}). This leads to a set of eigenvalues corresponding to the excitation energies.
    \item If properties of the excited states are needed, use the obtained $l$- and $r$-amplitudes after the diagonalization to construct the needed reduced density matrices (\ref{rdm1_ES}).
\end{enumerate}

\subsection{Experimentally Constrained Wavefunction Coupled Cluster approach (ECW-CC)}

We now wish to apply the coupled cluster formalism to the constrained wavefunction approach. Starting from the "pseudo" Schr\"odinger equation containing the experimental potential $V^{nm}_{\text{exp.}}$ (\ref{Sch_gene}), we repeat the steps from Sec. \hyperref[sec:III A]{III A}. $T$-equations (\ref{T_eq}) transform into  
\begin{equation}
\label{new_T_eq}
    \bra{\psi_\nu}\bar{H}+\bar{V}_{\text{exp.}}^{00}\ket{\psi_{\text{HF}}}=-\sum_{m \neq 0}\bra{\psi_\nu}\bar{V}_{\text{exp.}}^{0m}\hat{R}(m)\ket{\psi_{\text{HF}}}.
\end{equation}
The term on the right hand side leads to additional terms with respect to the traditional T-equations. As can be seen, the ground and excited states are coupled by the similarity transformed $\bar{V}_{\text{exp.}}^{0m}=e^{-\hat{T}}\hat{V}_{\text{exp.}}^{0m}e^{\hat{T}}$ potentials. Where $\bar{V}_{\text{exp.}}^{00}$ corresponds to the potential from the XCW method (involving ground state properties). These potentials contain the reduced density matrix (\ref{A through gamma}) used to predict experimental observations. According to equation (\ref{rdm1_ES}), they depend on the set of parameters used to describe the ground state ($t$- and $\lambda$-amplitudes) and the excited state ($r$- and $l$-amplitudes). In terms of the coupled cluster schemes, it means that the $T$-, $\Lambda$-, $R$- and $L$- equations become coupled and must be solved self-consistently. The expression (\ref{ground state energy}) transforms into
\begin{multline}
\label{new ground state energy}
    E_0=\bra{\psi_{\text{HF}}}\bar{H}-\bar{V}_{\text{exp.}}^{00}\ket{\psi_{\text{HF}}}\\-\sum_{m \neq 0}\bra{\psi_{\text{HF}}}\bar{V}_{\text{exp.}}^{0m}\hat{R}(m)\ket{\psi_{\text{HF}}}
\end{multline}
and loses its meaning as ground state energy. It  can be interpreted as a Lagrange multiplier, involved into the new $\Lambda$-equations
\begin{multline}
\label{new_Lam_eq}
    \bra{\psi_{\text{HF}}}(1+\hat{\Lambda})(\bar{H}-\bar{V}^{00}_{\text{exp.}})\ket{\psi_{\nu}}\\ 
-\sum_{m \neq 0} \bra{\psi_{\text{HF}}}\hat{L}(m)\bar{V}^{m0}_{\text{exp.}}\ket{\psi_{\nu}} = E_0\bra{\psi_{\text{HF}}}\hat{\Lambda}\ket{\psi_{\nu}}.
\end{multline}
Due to the experimental constraints, these equations become nonlinear with respect to the $\lambda$-parameters. Similarly to the new T-equations (\ref{new_T_eq}), the term inside the summation over $m$ corresponds to additional terms with respect to the standard CC-EOM equations.

The last step is to get EOM equations for the excited states. Following the recipe from Sec. \hyperref[sec: III B]{III B}, we start from the ansatz for $\ket{\Psi_n}$ (\ref{ES_CC_R}) and $\bra{\Psi_n}$ (\ref{ES_CC_L}).  Since the Experimentally Constrained Schr\"odinger equation (\ref{Sch_gene}) does not ensure strict orthogonality of the resulting states, we have to update the form of $\hat{L}(n)$ and add a term $l_0(n)$, that does not change the state
\begin{equation*}
    \hat{L}(n)=l_0(n)+\hat{L}_{1}(n)+\hat{L}_{2}(n)+...
\end{equation*}
In the traditional formalism, this term disappears because $0=\braket{\Psi_n|\Psi_0}=\braket{\psi_{\text{HF}}|\hat{L}|\psi_{\text{HF}}}=l_0(n)$. Inclusion of the experimental potentials in EOM method (\ref{EOM}) results in the new $R$-equations 
\begin{equation}
    \label{ECW_R}
\begin{split}
    \bra{\psi_{\text{HF}}}\bar{H}\hat{R}(n)\ket{\psi_{HF}}-\sum_m\bra{\psi_{\text{HF}}}\bar{V}^{nm}_{\text{exp.}}\hat{R}(m)\ket{\psi_{\text{HF}}} \\= E_{n}r_0(n),\!\!\\
\bra{\psi_{\nu}}\bar{H}\hat{R}(n)\ket{\psi_{\text{HF}}}-\sum_m\bra{\psi_{\nu}}\bar{V}^{nm}_{\text{exp.}}\hat{R}(m)\ket{\psi_{\text{HF}}} \\= E_{n}\bra{\psi_{\nu}}\hat{R}(n)\ket{\psi_{\text{HF}}}\!,\!
\end{split}
\end{equation}
and $L$-equations 

\begin{equation}
    \label{ECW_L}
\begin{split}
    \bra{\psi_{\text{HF}}}\hat{L}(n)\bar{H}\ket{\psi_{HF}}-\sum_m\bra{\psi_{\text{HF}}}\hat{L}(m)\bar{V}^{mn}_{\text{exp.}}\ket{\psi_{\text{HF}}} \\= E_{n}l_0(n),\!\!\\
\bra{\psi_{\text{HF}}}\hat{L}(n)\bar{H}\ket{\psi_{\nu}}-\sum_m\bra{\psi_{\text{HF}}}\hat{L}(m)\bar{V}^{mn}_{\text{exp.}}\ket{\psi_{\nu}} \\= E_{n}\bra{\psi_{\text{HF}}}\hat{L}(n)\ket{\psi_{\nu}}\!.\!
\end{split}
\end{equation}
Similarly to the $\Lambda$-equations, these are not linear anymore. Besides that, $L$-equations got extra equations for $l_0(n)$. 
To conduct the numerical analysis, we employ a method akin to that described in the previous subsection. First, all formal equations are reformulated into an algebraic form. When the observables derived from the experiment pertain to one- and two-particle properties—such as the dipole moment, transition energy, or structure factors—the Hamiltonian retains its dependence solely on at most two-particle operators. As a result, the algebraic equations for the numerical investigations are constructed from the same foundational elements as the conventional coupled-cluster (CC) equations. By utilizing the diagrammatic technique \cite{Crawford2007}, we express the terms in these equations directly in terms of the real $t$-, $\lambda$-, $r$- and $l$-amplitudes. 
Unfortunately, the equations are now interdependent, which prevents us from dividing the solving algorithm into distinct, independent steps. This means we can no longer start with the $T$-equations for the ground state and subsequently solve the $L$- and $R$-equations for the excited states.


Original $\Lambda$-, $L$- and $R$-equations (\ref{EOM}), (\ref{Lam_eq}) can be seen as an eigenvalue problem. Its solution gives directly spectrum and parametrization of all needed excited states. Now, $\Lambda$-, $L$- and $R$-equations (\ref{new_Lam_eq}), (\ref{ECW_R}), (\ref{ECW_L}) become non-linear and vary for different excited states, because of different experimental potentials (\ref{experimental_potential}). Instead of solving an eigenvalue problem, we treat different $\Lambda$-, $L$- and $R$-equations separately. 

To solve the resulting system of equations, we suggest the following iterative scheme. First, we define the ratio of theory and experiment by setting values of the Lagrange multipliers in the experimental constrain $Q[\Psi_0,\Psi_1,..., \Psi_N]$. The initial values for the $t$-, $\lambda$-, $r$- and $l$-amplitudes come from, for example, second order perturbation theory for the former and Koopman's excitation for the latter. This choice of initial guesses corresponds to the standard initial guess for CC and associated EOM implementation. In addition, we use the HF molecular orbitals as basis, allowing the use of the standard CC and EOM equations (with the additional terms described in the previous sections). At each step of the iterative process, we perform the following operations:
\begin{enumerate}
    \item One computes reduced one-particle density matrix elements (\ref{rdm1_ES}) using $t$-, $\lambda$-, $r$- and $l$-amplitudes from the previous steps.
    \item The resulting reduced density matrix can be used to determine the matrix elements of the experimental potential (\ref{experimental_potential}).
    \item At this point, the experimental potentials are known and can be treated as additional contributions to the matrix elements of the traditional Hamiltonian. The only difference is that this contribution differs for various states.
    \item Now, one solves $T$-equations (\ref{new_T_eq}) and determines (\ref{new ground state energy}) as if it was usual Coupled Cluster method.
    \item To finish the iteration step we solve the EOM equations. Since the experimental potential depends on the variables from the previous step, $\Lambda$-, $L$- and $R$-equations (\ref{new_Lam_eq}), (\ref{ECW_R}), (\ref{ECW_L}) are linear with respect to the variables from the current iteration.
    \item The described actions can be repeated until satisfactory convergence is achieved.
\end{enumerate}

\section{\label{sec:disc_persp} Discussion and perspective}


In this article, we present a comprehensive theoretical framework for a novel Experimental Constrained Wavefunction (ECW) method, alongside detailed guidelines for its future implementation using two commonly employed models—Hartree-Fock (HF) and Coupled Cluster (CC) Equation of Motion (EOM). Each of these models comes with distinct strengths and limitations.

The HF ansatz, although providing less accurately parameterized wavefunctions, requires significantly less computational effort compared to the more demanding CC implementation, making it an ideal starting point for initial numerical tests. Additionally, the Hartree-Fock method can be easily extended to density functional theory (DFT), offering flexibility in its application. An advantage of the HF ansatz is its complete independence in parameterizing ground and excited states, unlike the EOM method, which relies on using the CC ground state as a reference. Consequently, the EOM approach does not fully account for orbital relaxations. 

On the downside, the HF implementation's single-determinant ansatz is typically inadequate for capturing the diverse symmetries present in both ground and excited states. Furthermore, the unrestricted HF parametrization often results in spin contamination, which can compromise the accuracy of the results.

Like the XCW approach, the ECW technique encounters challenges when attempting to enhance the fit to experimental data by employing large Lagrange multipliers, which emphasize the experimental constraints in the variational problem (\ref{general minimization problem}). In the extreme case of infinitely large Lagrange multipliers, the theoretical component of the minimization problem becomes negligible, leading to an ill-posed problem. In practice, increasing the Lagrange multipliers typically results in a steeper gradient of the cost function in specific directions of the parameter space (which are unknown a priori), diminishing the prominence of other directions. Traditional gradient descent methods fail to account for this, leading to prolonged convergence cycles. This issue can potentially be mitigated by adopting a more sophisticated iterative scheme, such as introducing "friction" in the direction of the steeper descent.

Ideally, electronic wavefunctions should be optimized in tandem with molecular geometry. Similar to how the XCW method is coupled with the HAR approach, integrating geometry optimization into our framework is essential for practical relevance.

Looking ahead, we aspire for our robust theoretical framework to evolve into a comprehensive platform that leverages data from various complementary experimental techniques to construct a unified model of the underlying electronic structure. With the continual advancements in the precision of experimental methods and the routine combination of multiple techniques to achieve a deeper understanding of complex systems, the ability to enhance or refine theoretical models using such experimental data is a future topic of potentially significant impact.

\begin{acknowledgments}
We acknowledge the financial support of Grant-No.~HIDSS-0002 DASHH (Data Science in Hamburg-Helmholtz Graduate School for the Structure of Matter) and the BMBF project MeSoX (Project No. 05K19GUE). This research is partially supported by the Deutsches Elektronen-Synchrotron (DESY) Strategy Fund. We are grateful to Prof. Dr. Dylan Jayatilaka, Prof. Dr. Alessandro Genoni, Dr. Simon Grabowsky, Prof. Dr. Christina Brandt and Dr. Thi Bich Tram Do for fruitful discussions.
\end{acknowledgments}

\bibliography{anotherlibrary,bibliography}

\appendix

\section{Explicit form of Fock matrix elements\,$\bar{f}^{(n)}_{\mu\nu}$}
\label{technical}

Similarly to equation (\ref{modified fock}), we split the Fock matrix $ \bar{f}_{\mu\nu}^{(n)}$ into three contributions
\begin{subequations}
\label{eq: technical}
\begin{equation}
    \bar{f}_{\mu\nu}^{(n)}= f_{\mu\nu}^{(n)}+v_{\mu\nu}^{(n)}+\sum_{m\neq n} v_{\mu\nu}^{(n,m)}.
\end{equation}
The first term is the matrix element  $f_{\mu\nu}^{(n)}$ of the traditional Fock operator for state $n$ in the common basis set. $v_{ia}^{(n)}$ is the contribution from the experimental data related solely to state $n$
\begin{widetext}
\begin{equation}
    v_{\mu\nu}^{(n)}=\sum_{\sigma,\delta}\left(V^{nn}_{\mu\sigma}\gamma^{nn}_{\sigma\delta}S_{\delta\nu}+S_{\mu\sigma}\gamma^{nn}_{\sigma\delta}V^{nn}_{\delta\nu}-2S_{\mu\sigma}\sum_{\beta,\rho}\gamma^{nn}_{\sigma\beta}V^{nn}_{\beta\delta}\gamma^{nn}_{\delta\rho}S_{\rho\nu}\right).
\end{equation}
Finally, $v_{\mu\nu}^{(n,m)}$ represents the experimental data related to pairs of different states
\begin{multline}
    v_{\mu\nu}^{(n,m)}=\sum_{\sigma,\rho,\kappa,\delta',\delta}\left[\delta_{\mu\sigma}-\sum_{\sigma'}S_{\mu\sigma'}\gamma^{nn}_{\sigma'\sigma}\right]\left[ V^{nm}_{\sigma\rho}\gamma^{mn}_{\rho\kappa}+\sum_{\mu',\nu'}S_{\sigma \rho}\gamma^{mn}_{\rho\mu'\kappa\nu'}V^{nm}_{\nu'\mu'}\right]S_{\kappa\delta}\gamma^{nn}_{\delta \delta'}S_{\delta'\nu}\\+\sum_{\sigma,\rho,\kappa,\delta',\delta} S_{\mu\delta'}\gamma^{nn}_{\delta'\delta }S_{\delta\kappa}\left[ \gamma^{nm}_{\kappa\rho}V^{mn}_{\rho\sigma}+\sum_{\mu',\nu'}V^{mn}_{\mu'\nu'}\gamma^{nm}_{\kappa\nu'\rho\mu'}S_{\rho\sigma }\right]\left[\delta_{\sigma\nu}-\sum_{\sigma'}\gamma^{nn}_{\sigma\sigma'}S_{\sigma'\nu}\right].
\end{multline}
\end{widetext}
\end{subequations}

\end{document}